\documentclass[floatfix,twocolumn,showpacs,preprintnumbers,amsmath,amssymb,pra,superscriptaddress,longbibliography]{revtex4-1}
\usepackage{color}
\usepackage[usenames,dvipsnames,svgnames,table]{xcolor}
\usepackage[colorlinks=true,linkcolor=blue,urlcolor=blue,citecolor=blue]{hyperref}
\usepackage{mathtools}
\usepackage{graphicx}
\usepackage{dcolumn}
\usepackage{array}
\usepackage{lipsum}
\usepackage{bm}
\usepackage{subfigure}
\usepackage{amssymb}
\usepackage{multirow}
\usepackage{tabularx}
\usepackage{amsmath}
\usepackage{braket}
\usepackage{csquotes}
\graphicspath{{plots/}}
 \usepackage{lipsum}
\usepackage{mathrsfs}
\usepackage{MnSymbol}

\newcommand{\beq}{\begin{equation}}
\newcommand{\eeq}{\end{equation}}
\newcommand{\bea}{\begin{eqnarray}}
\newcommand{\eea}{\end{eqnarray}}

\begin{document}
\title{Ultrahigh Resolution X-ray Thomson Scattering Measurements\\ at the European XFEL}

\author{Thomas Gawne}
\email{t.gawne@hzdr.de}

\affiliation{Center for Advanced Systems Understanding (CASUS), D-02826 G\"orlitz, Germany}
\affiliation{Helmholtz-Zentrum Dresden-Rossendorf (HZDR), D-01328 Dresden, Germany}

\author{Zhandos A.~Moldabekov}

\affiliation{Center for Advanced Systems Understanding (CASUS), D-02826 G\"orlitz, Germany}
\affiliation{Helmholtz-Zentrum Dresden-Rossendorf (HZDR), D-01328 Dresden, Germany}

\author{Oliver~S.~Humphries}
\affiliation{European XFEL, D-22869 Schenefeld, Germany}



\author{Karen Appel}
\affiliation{European XFEL, D-22869 Schenefeld, Germany}

\author{Carsten Baehtz}
\affiliation{Helmholtz-Zentrum Dresden-Rossendorf (HZDR), D-01328 Dresden, Germany}

\author{Victorien Bouffetier}
\affiliation{European XFEL, D-22869 Schenefeld, Germany}

\author{Erik Brambrink}
\affiliation{European XFEL, D-22869 Schenefeld, Germany}

\author{Attila Cangi}
\affiliation{Center for Advanced Systems Understanding (CASUS), D-02826 G\"orlitz, Germany}
\affiliation{Helmholtz-Zentrum Dresden-Rossendorf (HZDR), D-01328 Dresden, Germany}

\author{Sebastian G\"ode}
\affiliation{European XFEL, D-22869 Schenefeld, Germany}

\author{Zuzana Kon\^opkov\'a}
\affiliation{European XFEL, D-22869 Schenefeld, Germany}

\author{Mikako Makita}
\affiliation{European XFEL, D-22869 Schenefeld, Germany}

\author{Mikhail Mishchenko}
\affiliation{European XFEL, D-22869 Schenefeld, Germany}

\author{Motoaki Nakatsutsumi}
\affiliation{European XFEL, D-22869 Schenefeld, Germany}

\author{Kushal Ramakrishna}
\affiliation{Center for Advanced Systems Understanding (CASUS), D-02826 G\"orlitz, Germany}
\affiliation{Helmholtz-Zentrum Dresden-Rossendorf (HZDR), D-01328 Dresden, Germany}

\author{Lisa Randolph}
\affiliation{European XFEL, D-22869 Schenefeld, Germany}

\author{Sebastian Schwalbe}
\affiliation{Center for Advanced Systems Understanding (CASUS), D-02826 G\"orlitz, Germany}
\affiliation{Helmholtz-Zentrum Dresden-Rossendorf (HZDR), D-01328 Dresden, Germany}

\author{Jan Vorberger}
\affiliation{Helmholtz-Zentrum Dresden-Rossendorf (HZDR), D-01328 Dresden, Germany}

\author{Lennart Wollenweber}
\affiliation{European XFEL, D-22869 Schenefeld, Germany}

\author{Ulf Zastrau}
\affiliation{European XFEL, D-22869 Schenefeld, Germany}

\author{Tobias Dornheim}
\email{t.dornheim@hzdr.de}
\affiliation{Center for Advanced Systems Understanding (CASUS), D-02826 G\"orlitz, Germany}
\affiliation{Helmholtz-Zentrum Dresden-Rossendorf (HZDR), D-01328 Dresden, Germany}

\author{Thomas~R.~Preston}
\email{thomas.preston@xfel.eu}
\affiliation{European XFEL, D-22869 Schenefeld, Germany}

\begin{abstract}
   Using a novel ultrahigh resolution ($\Delta E\sim0.1\,$eV) setup to measure electronic features in x-ray Thomson scattering (XRTS) experiments at the European XFEL in Germany, we have studied the collective plasmon excitation in aluminium at ambient conditions, which we can measure very accurately even at low momentum transfers. As a result, we can resolve previously reported discrepancies between \emph{ab initio} time-dependent density functional theory simulations and experimental observations. The demonstrated capability for high-resolution XRTS measurements will be a game changer for the diagnosis of experiments with matter under extreme densities, temperatures, and pressures, and unlock the full potential of state-of-the-art x-ray free electron laser (XFEL) facilities to study planetary interior conditions, to understand inertial confinement fusion applications, and for material science and discovery.
\end{abstract}

\maketitle


The x-ray Thomson scattering (XRTS) technique~\cite{sheffield2010plasma} has emerged as a powerful method of diagnosing matter. By probing the electronic dynamic structure factor $S_{ee}(q,E)$, where $q$ and $E$ are the change in the momentum and energy of the scattered photon, it is capable of giving detailed insights into the microphysics of the probed sample~\cite{siegfried_review,Gregori_PRE_2003,Dornheim_review}.
This capability is particularly important for experiments with matter under extreme densities, temperatures, and pressures~\cite{wdm_book,drake2018high,siegfried_review}, as they occur e.g.~in astrophysical objects~\cite{Benuzzi_Mounaix_2014,becker,Kritcher2020}, inertial confinement fusion applications~\cite{hu_ICF,Betti2016}, and for material science and materials discovery~\cite{Kraus2016,Kraus2017,Lazicki2021}.
Here, the combination of the extreme conditions with the highly transient nature of the generated extreme states in the laboratory~\cite{Vorberger_PLA_2024} renders the unambiguous diagnosis of plasma conditions challenging.

Since the first observation of plasmons in warm dense beryllium~\cite{Glenzer_PRL_2007}, a number of major developments have helped to establish XRTS as a reliable method for the study of materials over a vast range of densities and temperatures. This includes the demonstration of ultrabright seeded x-ray free-electron laser (XFEL) beams~\cite{Fletcher2015}, the utilization of XRTS for the detection and quantification of miscibility~\cite{Frydrych2020}, and the resolution of ion acoustic modes in single-crystal diamond~\cite{Descamps_SR_2020}.


The measured XRTS intensity is given by~\cite{siegfried_review,sheffield2010plasma,Dornheim_T2_2022}
\begin{eqnarray}\label{eq:convolution}
    I(q,E_s) = S_{ee}(q,E_0-E_s) \circledast R(E_s)\ , 
\end{eqnarray}
i.e., as a convolution of $S_{ee}(q,E)$ with the combined source-and-instrument function $R(E_s)$, where $E_0$ and $E_s$ denote the beam energy and the energy of the scattered photon. In practice, deconvolving Eq.~(\ref{eq:convolution}) is generally numerically unstable; the 
width of $R(E_s)$ thus limits
the capability of XRTS to resolve electronic features such as sharp plasmon excitations.
This strongly hampers the model-free diagnostics
of parameters such as the temperature~\cite{Dornheim_T_2022,Dornheim_T2_2022} that are important for equation-of-state measurements~\cite{Falk_PRL_2014,Tilo_Nature_2023}, and poses a serious obstacle for the benchmarking of theoretical models against experimental observations~\cite{Cazzaniga_PRB_2011,Ramakrishna_PRB_2021}.


\begin{figure}
    \centering
    \includegraphics[width=0.475\textwidth,keepaspectratio]{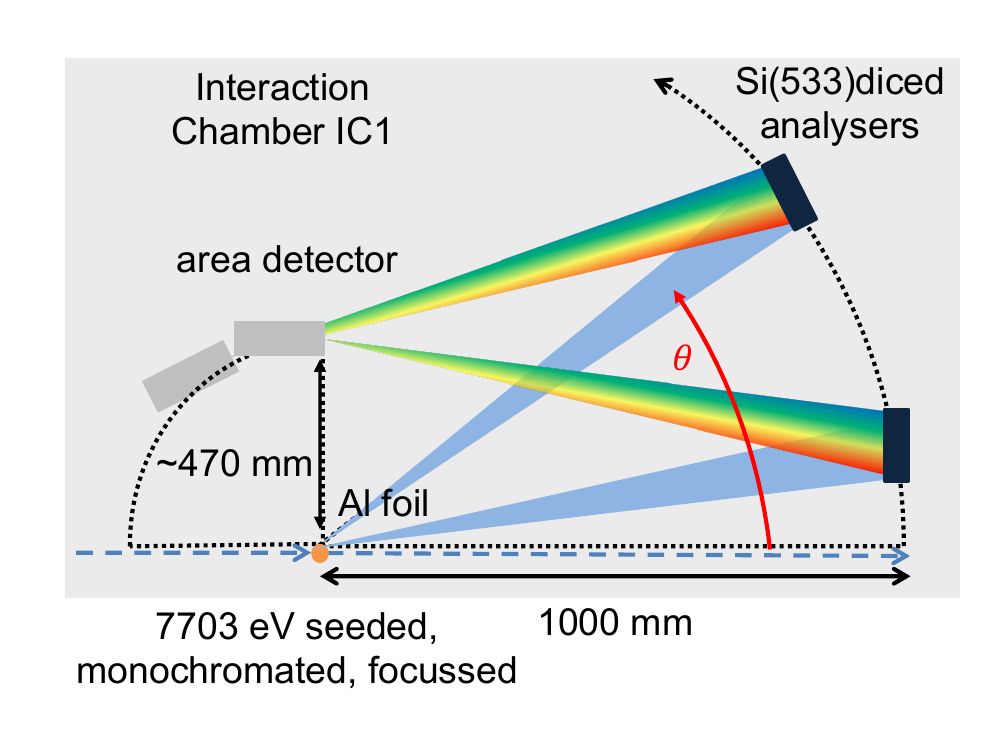}
    \caption{Schematic illustration of the set-up in Interaction Chamber 1 (IC1) modified from~\citet{Wollenweber_RSI_2021}. The x-ray beam is seeded at 7703\,eV in the SASE2 undulator including a SASE pedestal. The beam is passed through a four-bounce Si (111) monochromator to remove this SASE pedestal before being focused onto an Al foil. The x-rays are scattered, collected and focused by a spherically bent diced analyzer crystal onto a Jungfrau detector with asymmetric pixels. Both the analyzer and detector are mounted onto curved rails to vary the scattering angle $\theta$. The detector is shown in two configurations: at $13.6^\circ$  corresponding to $(\pi/2 - \theta_B)$ the Bragg angle with the detector plane directly above the sample; and the highest angle. The scattering angle can be freely set between $3.6^\circ$ and $25.6^\circ$ in this study. The combination of the monochromated beam and ultra high resolution spectrometer allows for the high-fidelity measurements of electronic structure.}
    \label{fig:scheme}
\end{figure}

In this Letter, we present measurements of the plasmon in Al with an unprecedented resolution of $\Delta E\sim0.1\,$eV that allows us to resolve electronic features from XRTS measurements and reconcile discrepancies in modelling. Our measurements of Al at ambient conditions, made with a new set-up at the European XFEL in Germany~\cite{Tschentscher_2017}, using a seeded and monochromated beam, are in excellent agreement with previous electron energy loss spectroscopy (EELS) measurements~\cite{Krane_JPF_1978,sprosser1989aluminum}.
In addition, the high resolution of our results allows us to unambiguously benchmark time-dependent density functional theory (TDDFT)~\cite{Runge_PRL_1984}
calculations, and to resolve discrepancies reported in previous works~\cite{Cazzaniga_PRB_2011,Ramakrishna_PRB_2021}.

This new capability will be of paramount importance for XRTS diagnostics: first and foremost, it allows for the model-free interpretation of the measured spectra even for comparably moderate temperatures of $T\sim1\,$eV, whereas previous experimental set-ups were limited to $T>10\,$eV~\cite{Dornheim_T_2022,Dornheim_T2_2022}. This will unlock the full potential of modern XFEL facilities for material science and discovery~\cite{Kraus2016,Kraus2017,Lazicki2021}, for the characterization of the initial phase of the compression path of the fuel capsule in laser fusion experiments, and for the study of planetary interior conditions~\cite{Benuzzi_Mounaix_2014}. We note that, despite its comparable resolution, EELS generally cannot be used to diagnose experiments in this regime due to its stringent requirements for thin targets and long measurement times~\cite{brydson2020electron}. Furthermore, to temporally resolve XRTS in WDM requires a femtosecond x-ray free electron laser such as that at the European XFEL. Due to the short femtosecond timescales of FELs we can isochorically heat with the FEL beam to temperatures in the WDM matter regime~\cite{Witte_PRL_2017}. If defocussed, the FEL can also be used as a fs-probe in the event of another driver being used in a pump-probe setup, enabling studies of highly transient states which cannot be performed elsewhere.

In addition, we have demonstrated the capability of our XRTS setup to produce rigorous benchmark data for first-principles theoretical methods, which is indispensable for the development of new methodologies to simulate extreme states of matter~\cite{wdm_book,new_POP,Dornheim_review}.
Finally, the detailed resolution of electronic features of different materials constitutes an important end in itself, and promises novel insights into the behaviour of matter across various density and temperature regimes.

\textbf{Experimental set-up.} The experiment was performed at the HED instrument~\cite{Zastrau2021} of the European XFEL, Germany. The XFEL beam was self-seeded to an energy of $E_0\sim7703$~eV, and then passed through a four-bounce silicon $(111)$ monochromator with an acceptance range of 0.8\,eV to remove the underlying self-amplified spontaneous emission (SASE) pedestal from the beam. 
The beam was focused onto a 50$\mu$m thick Al foil using a set of Be compound refractive lenses~\cite{Zastrau2021} located 9\,m upstream to a spot of order 10$\mu$m. The full width at half maximum (FWHM) of the beam incident on the sample was measured to be $\Delta E\sim 450-530$ meV, corresponding to a spectral bandwidth of $\Delta E/E\sim5.8-6.9~\times 10^{-5}$, as measured from the quasi-elastic scattering feature.
An upstream gas monitor measured initial beam energies of $146-310~\mu$J, however transmission through the beam line optics ($\sim70~\%$) and after removal of the large SASE pedestal in the monochromator reduced the energy on target to $15.5-21.8~\mu$J ($\sim7~\%$ transmission), as measured on a second gas monitor before the target. Removing the SASE pedestal was critical in this experiment since the seeding performance was unsatisfactory and $\sim85\%$ of the fluence was not in the seed. Furthermore, due to the convolution of Eq.~\ref{eq:convolution} the brighter and broader pedestal would blur our measured XRTS signal. 

The scattered x-rays were collected by a spherically-bent Si $(533)$ diced crystal analyser (DCA) previously employed to measure phonons~\cite{Descamps_SR_2020} and described in more detail by Ref.~\cite{Wollenweber_RSI_2021}. However, in contrast to Ref.~\cite{Wollenweber_RSI_2021} where scattering was measured in an spectral range of just 0.3~eV, here we measure the signal up to energy losses of 40~eV. The incident photons collected by the DCA are dispersed and focused onto a Jungfrau detector~\cite{Mozzanica2018} employing asymmetric pixels, with a size of 25\,$\mu$m in the dispersive direction and 225\,$\mu$m in the non-dispersive direction. The detector is single-photon sensitive with an excellent noise of $150$\,eV in the highest gain stage (or 0.02 photons at 7700\,eV)~\cite{Mozzanica2016}. Individual images can be thresholded to eliminate electronic noise and detect only single photon events giving a noise level well below the Poisson statistical noise for single photon detection events~\cite{Mozzanica2016}. It is the combination of monochromating the incident beam and collection of the scattered photons on the ultra-high resolution DCA spectrometer that allows for high fidelity measurements of the XRTS signal on electronic energy scales.

The primary source of uncertainty in the plasmon measurement is due to the finite angular coverage of the DCA. This results in so-called $q$-broadening, where the inelastic signal measured is integrated over a range of scattering vectors. To narrow the angular coverage of the DCA, a horizontal slit mask made from Al was placed on the DCA, which reduced its coverage to $\pm 1.4^\circ$, or a $q$ range of $\pm0.095~{\rm \AA^{-1}}$ to $\pm0.098~{\rm \AA^{-1}}$. This slit width was chosen as a compromise between reducing the $q$-broadening and maximising the reflectivity of the DCA through the number of reflecting dice~\cite{Wollenweber_RSI_2021}. For our measurement the DCA has uniform reflectivity in our $q$-window, and the calculated standard deviation of our $q$ measurement is $\sim \pm 0.03$\,\AA$^{-1}$.

The DCA reflects a spectral window of 3.5\,eV at the Bragg angles used in this experiment, so it must be scanned in Bragg angle to collect spectra at different energies.
To mitigate the drop off in reflectivity in the wings of the spectra~\cite{Wollenweber_RSI_2021} the DCA was conservatively scanned in small steps so that the spectral windows overlap. Relaxing this condition, the plasmon feature could be collected in only two to three steps since the window is so wide, substantially reducing the number of shots required to collect the plasmon. An estimate of the ${\rm SNR}=\sqrt{NI}$ (signal-to-noise ratio), where $I$ is the integrated intensity in photons/shot and $N$ the number of shots. We estimate that a total of $10^3-10^4$ shots should be sufficient for a ${\rm SNR}\sim6-28$ at the plasmon feature, which requires a collection time of $<20$\,minutes at 10\,Hz; fewer shots are required with higher flux. Better seeding performance could allow the omission of the monochromator, further increasing the intensities achieved at the sample.

The recorded spectral window  was calibrated using Co K$_{\beta}$ emission and the position of the quasi-elastic scattering. This was established accurately by tuning the seeded beam energy to cross the Co K-edge and observing when the intensity of the elastically scattered photons plateaued relative to the Co K-shell emission recorded by a spectrometer~\cite{Preston2020} viewing the upstream side of a Co foil. The incident photon beam was determined from this method to have a central photon energy of $E_0=7703.21$~eV for this study on Al, and the energy dispersion was determined to be 22.54~meV/pixel.

\begin{figure}
    \centering
    \includegraphics[width=0.475\textwidth,keepaspectratio]{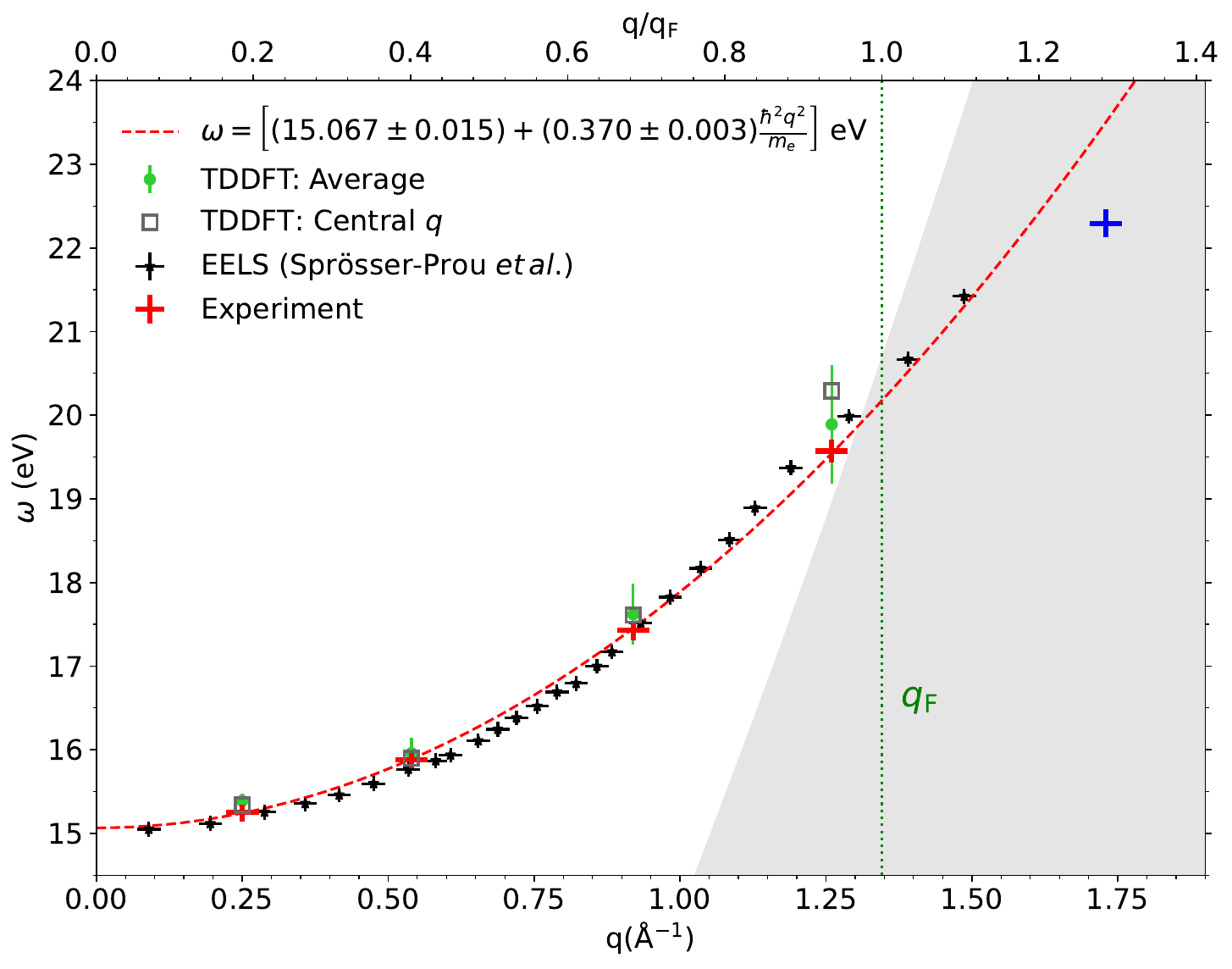}
    \caption{ Plasmon dispersion of aluminium. Crosses: experimental peak position and associated uncertainty, with red below the pair continuum, and blue in the pair continuum; red dashed line: the experimental data fitted to the dispersion of Eq.~\ref{eq:BG}; black stars: previous EELS measurements of the Al plasmon by Spr\"osser-Prou \textit{et al.}~\cite{sprosser1989aluminum}; green circles: TDDFT average plasmon position, with the error bars indicating the standard deviation of the peak position in the $q$ range; grey squares: TDDFT plasmon position only at the central $q$ value. The shaded grey area indicates the pair continuum~\cite{quantum_theory}, and the dotted vertical line the Fermi wavenumber $q_\textnormal{F}$ in bulk Al with a face-centered cubic (fcc) lattice \cite{moldabekov2024excitation}.
    }
   \label{fig:3}
\end{figure}

To measure the plasmon at a range of individual scattering vectors, the DCA setup was moved to different scattering angles between 3.6$^\circ$ to 25.6$^\circ$, corresponding to $q=0.245-1.730$~\AA$^{-1}$. 
This allows 
the dynamic structure factor of
Al to be probed
from the collective regime up into the electron-hole pair continuum, cf.~Fig.~\ref{fig:3}.

\begin{figure}
    \centering
    \includegraphics[width=0.475\textwidth,keepaspectratio]{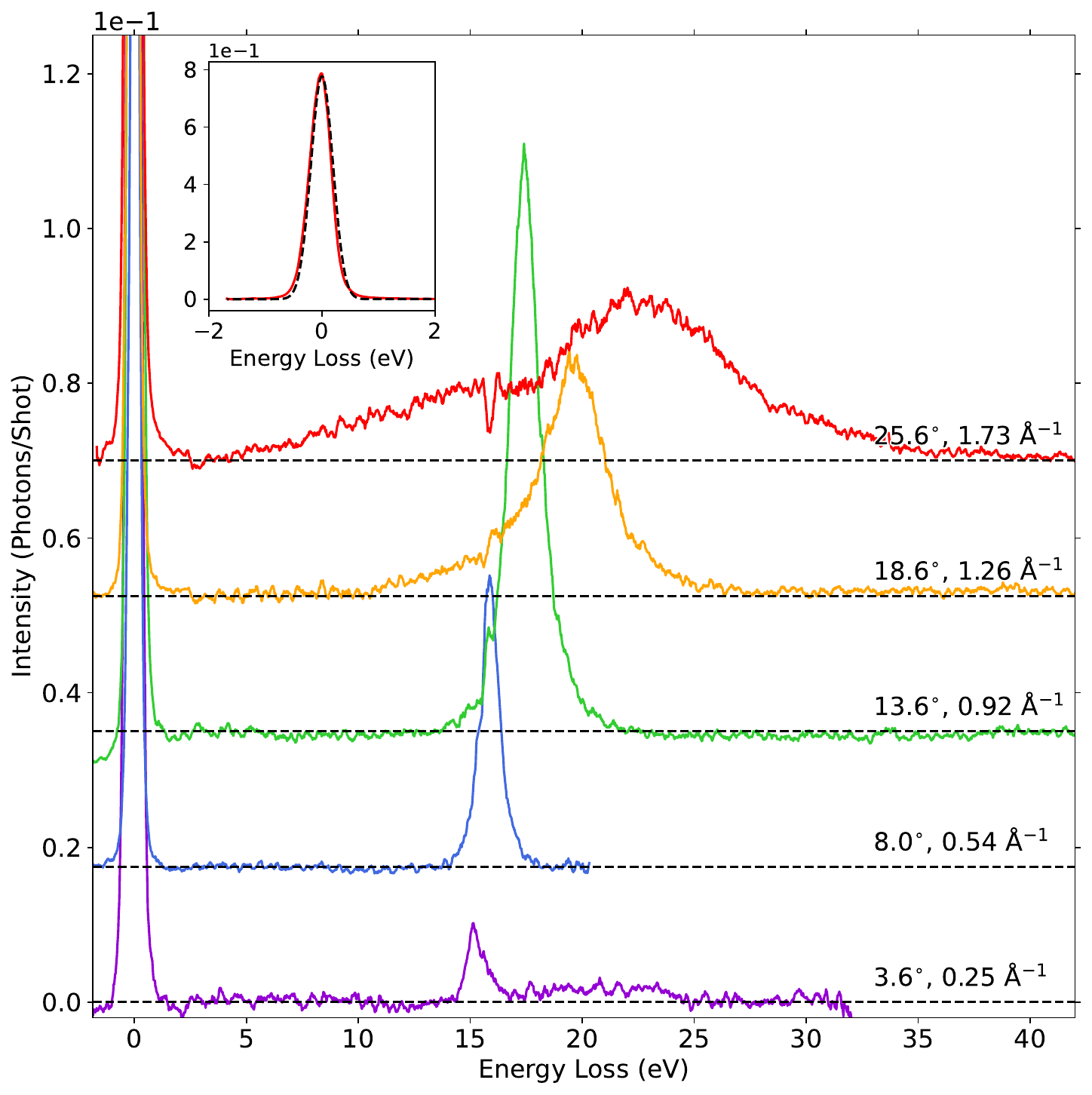}
    \caption{Measured XRTS intensity for five different wavenumbers as a function of the photon energy loss $E=E_0-E_s$ in units of integrated intensity in photons/shot. The curves are offset vertically for clarity and show the variation in position, intensity, and shape of the plasmon in aluminium. Further details on the collection of the spectra are provided in the table in the Supplemental material. Inset: example spectrum of the narrow quasi-elastic scattering (red) for the highest wavenumber, and a Gaussian fit with $\sigma=0.19$~eV (black dashed). The intensity of the quasi-elastic scattering uses the same scale as the main figure.
    }
   \label{fig:2}
\end{figure}

\textbf{Results.} In Fig.~\ref{fig:2}, we display the measured XRTS intensity as a function of the photon energy loss $E=E_0-E_s$.
The inset shows the quasi-elastic feature around $E=E_0$ corresponding to the combined source and instrument function $R(E_s)$; it is approximately Gaussian with a full width at half maximum of $\Delta E=0.46\,$eV ($\sigma=0.2\,$eV). The horizontal dashed lines indicate the zero intensity base lines for the five considered wavenumbers $q\in[0.25,1.73]\,$\AA$^{-1}$, and the corresponding intensities are shown as the coloured curves. Remarkably, the effect of the convolution of $S_{ee}(q,E)$ with $R(E_s)$ [cf.~Eq.~(\ref{eq:convolution})] can be neglected in practice due to the high resolution of our setup. At the same time, we are able to accurately resolve the plasmon down to below $20\%$ of the Fermi wave number due to the high brilliance of the XFEL, and its excellent repetition rate.  
This remarkable performance might even open up new possibilities to connect XRTS with the estimation of optical properties such as the dielectric function, opacity, and electrical conductivity, which are defined for $q\to0$~\cite{Witte_PRL_2017,Sperling_PRL_2015}.

The resulting dispersion relation of aluminium is shown in Fig.~\ref{fig:3}, where we indicate our measurement of the plasmon position as a function of $q$.
The finite size of the DCA covers a $q$-range of $\sim \pm 0.1$\,\AA$^{-1}$: to be clear, this does not represent an uncertainty in central $q$ value as the angular position of the DCA was initially measured and then carefully positioned using motors; instead, it represents the range of values of $q$ over which the XRTS spectrum is averaged.
As the number of die in the DCA is uniform in $q$, the $q$-uncertainty plotted in Fig.~\ref{fig:3} are the standard deviation for a uniform distribution, which gives an accuracy of $\sim \pm 0.03$\,\AA$^{-1}$, almost exactly the same as previous EELS measurements~\cite{sprosser1989aluminum}.
The accuracy in the plasmon shift is mainly due to the calibration of the energy axis, which is explained in more detail in the Supplemental Material~\cite{supplement}.
Identified plasmon peaks are fitted with the expected quadratic dispersion relation of the form:
\begin{equation}\label{eq:BG}
    \omega = \omega_p + \alpha \frac{\hbar^2 q^2}{m_e} \, ,
\end{equation} following the familiar Bohm-Gross relation~\cite{Bohm_Gross}, where $\hbar$ is the reduced Planck constant and $m_e$ is the electron mass. We determine a pre-factor of $\alpha=0.370\pm0.003$ and plasma frequency of $\omega_p=15.067\pm0.015\,$eV where the error is the standard deviation calculated using the total least squares, accounting for the uncertainties in $q$ and the plasmon shift.
We find excellent agreement with previous EELS measurements~\cite{sprosser1989aluminum,Krane_JPF_1978}, which further substantiates the high quality of our results, and indeed improve on the accuracy in the scaling parameter $\alpha$.
Above the electron--hole pair continuum~\cite{quantum_theory}, where the plasmon decays into a multitude of excitations due to Landau damping, the position of the maximum in $S_{ee}(q,E)$ deviates from Eq.~(\ref{eq:BG}), and the experimental spectrum attains a broad, nontrivial shape (see the curve in Fig.~\ref{fig:2} for highest $q$).

\begin{figure}
    \centering
    \includegraphics[width=0.475\textwidth,keepaspectratio]{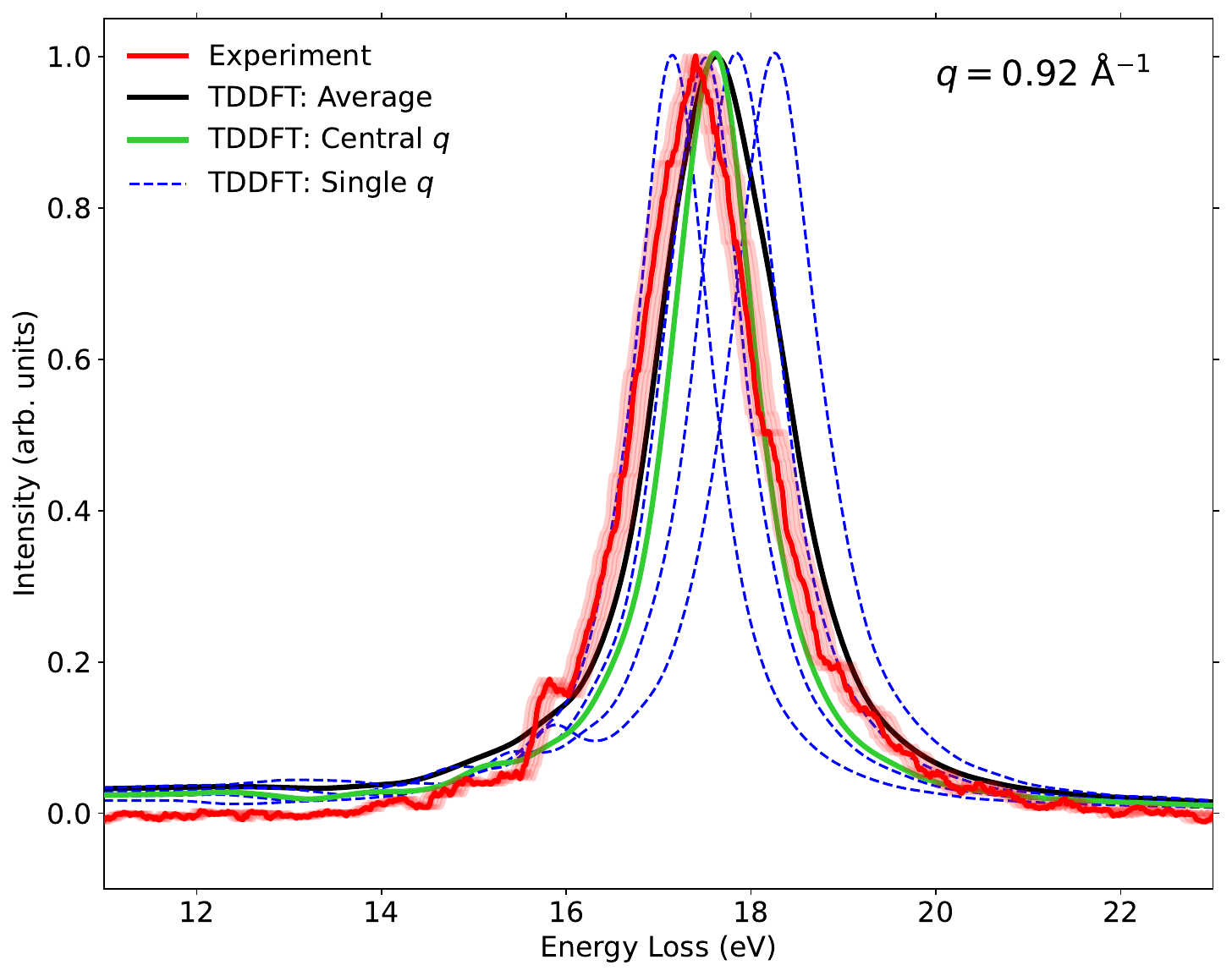}
    \caption{ Using high-resolution XRTS measurements (solid red) to benchmark first-principles TDDFT simulations at $q_0=0.92\,$\AA$^{-1}$. The solid green curve is from a single TDDFT simulation at $q_0=0.932\,$\AA$^{-1}$, the closest our simulation box size allows us to get to the central $q$ value. The blue dashed curves have been obtained from single TDDFT simulations at different $q$-values in the experimental range: 0.858, 0.914, 0.972, 1.029\,\AA$^{-1}$. The solid black line has been averaged over the 5 individual TDDFT results with equal weighting.
    The red area indicates the experimental uncertainty in the plasmon position.
    }
   \label{fig:4}
\end{figure}

Let us next utilize our new high-resolution XRTS data to assess the accuracy of \emph{ab initio} time-dependent density functional theory (TDDFT) simulations~\cite{ullrich2012time,Cazzaniga_PRB_2011,Ramakrishna_PRB_2021}. As a first step, we carried out linear-response TDDFT calculations employing the adiabatic local density approximation (ALDA) for the four lowest $q$-values; see Ref.~\cite{supplement} for technical details.
The corresponding peak positions only considering the central $q$ are included in Fig.~\ref{fig:3} and are in reasonable agreement with the experimental data points for the three smallest probed wavenumbers, but significantly overestimate the true plasmon position for $q=1.3\,$\AA$^{-1}$.
To get additional insights into the quality of the TDDFT simulations, we show a detailed comparison of the latter with the experimental intensity in Fig.~\ref{fig:4} for $q=0.92\,$\AA$^{-1}$, clearly demonstrating that TDDFT underestimates the peak width despite reproducing the peak position with good accuracy.

While the attribution of such deviations between experiment and simulation to a systematic error due to the employed approximation for the exchange--correlation functional or exchange--correlation kernel is tempting, other explanations have to be considered. First, it is noted that the earlier XRTS experiments had substantially broader source-and-instrument functions~\cite{Cazzaniga_PRB_2011}, which might obscure the true origin of an observed deviation.
The ultrahigh resolution achieved here resolves this conundrum and rules out any significant effects due to $R(E_s)$.
Second, the finite angular coverage of the employed DCAs implies that the experimental spectra are effectively averaged over a finite interval of $q$-values.
To rigorously take into account this effect, we have carried out a series of TDDFT simulations for a uniformly distributed set of wavenumbers $q\in[-\Delta q+q_0,\Delta q+q_0]$, and the results are included in Fig.~\ref{fig:4} (single $q$).
Interestingly, such slight changes in the wavenumber significantly affect the peak position, width, shape, and symmetry of the generated XRTS spectra.
As our best prediction of the measured XRTS intensity, these generated spectra are averaged over the $q$-range with equal weight~\cite{supplement}, providing substantially improved agreement with the experimental observation both with respect to its form and position (see average in Fig.~\ref{fig:4}).
While the TDDFT wings appear slightly higher than the measured curve, they are nevertheless within the experimental uncertainty in the intensity.
The same holds for the other three considered wavenumbers, which are shown in the Supplemental Material~\cite{supplement}.
In this way, our novel setup has allowed observed differences between TDDFT simulations and XRTS experiments to be reconciled, which we attribute to the mixing of wavenumbers for the present case.
For completeness, we have included the properly $q$-averaged peak positions in Fig.~\ref{fig:3} as well as an indication of the variance of the TDDFT results within the appropriate $q$-window.
The effect of averaging is small for $q\lesssim1\,$\AA$^{-1}$, but leads to a substantial improvement for the second largest wavenumber: here the importance of averaging correctly over the $q$-range is clearly demonstrated.


\textbf{Conclusion.} In this Letter, we have presented new ultrahigh resolution XRTS measurements of electronic features obtained at the European XFEL. Our setup works over a broad range of wavenumbers and can be implemented in both forward and backward scattering geometries in future experiments. 
As a practical example, the plasmon in aluminium at ambient conditions was studied, which was resolved with high accuracy even at very small scattering angles. This opens up the enticing opportunity to connect XRTS measurements with the estimation of optical properties such as the electrical conductivity.

We are convinced that our work opens up a variety of possibilities for important future research. First, resolving the electronic structure of different materials with high resolution is important in its own right and will give new insights into physical effects such as the predicted roton-type excitation in low-density hydrogen~\cite{Hamann_PRR_2023,Dornheim_Nature_2022}, or very recently reported thermal features in isochorically heated materials~\cite{moldabekov2024excitation, moldabekov2024ultrafast}. Recent performance on the SASE2 undulator~\cite{Liu2023} has demonstrated a spectral density of 1\,mJ/eV in the seeded beam which would yield fluences of $10^{11}$\,photons/pulse delivered at 10\,Hz. Whilst comparable to synchrotrons, this high fluence can be delivered at femtosecond timescales (typical FWHM of 25\,fs) and focussed into a small volume yielding intensities per pulse $\sim10^{17}$\,W/cm$^2$, heating the electronic system faster than the electron-ion coupling time, which is more than sufficient to heat into the WDM regime~\cite{Witte_PRL_2017}. In addition, the narrow width of $R(E_s)$ facilitates the rigorous benchmarking of theoretical models~\cite{Mo_PRB_2020,Zhang_2024} and first-principles simulations~\cite{dynamic2,Dornheim_review} such as the exchange--correlation functional and exchange--correlation kernel in TDDFT calculations~\cite{Moldabekov_JCTC_2023,Schoerner_PRE_2023,Cazzaniga_PRB_2011,Ramakrishna_PRB_2021}.

A particularly important field of application is given by the diagnostics of experiments with matter under extreme conditions.
Specifically, a narrow source-and-instrument function is key to extend the model-free interpretation of XRTS spectra to temperatures of the order of $T\sim1\,$eV, whereas previous efforts where restricted to $T\gtrsim10\,$eV~\cite{Dornheim_T_2022,Dornheim_T2_2022}.
This development promises to unlock the full capability of modern XFEL facilities such as the European XFEL in Germany~\cite{Tschentscher_2017} and the Linac Coherent Light Source (LCLS) in the USA~\cite{LCLS_2016} to study matter at planetary interior conditions~\cite{Benuzzi_Mounaix_2014}, to diagnose material science and material discover applications, and to characterize states that occur on the initial stage of the compression path of the fuel capsule in inertial confinement fusion experiments.

\begin{acknowledgements}
We acknowledge the European XFEL in Schenefeld, Germany, for provision of X-ray free-electron laser beamtime at the Scientific Instrument HED (High Energy Density Science) under proposal number 3777 and would like to thank the staff for their assistance. The authors are grateful to the HIBEF user consortium for the provision of instrumentation and staff that enabled this experiment. The original datasets can be found here and are available upon reasonable request: doi:10.22003/XFEL.EU-DATA-003777-00.

This work was partially supported by the Center for Advanced Systems Understanding (CASUS) which is financed by Germany’s Federal Ministry of Education and Research (BMBF) and by the Saxon state government out of the State budget approved by the Saxon State Parliament.
This work has received funding from the European Union's Just Transition Fund (JTF) within the project \emph{R\"ontgenlaser-Optimierung der Laserfusion} (ROLF), contract number 5086999001, co-financed by the Saxon state government out of the State budget approved by the Saxon State Parliament.
This work has received funding from the European Research Council (ERC) under the European Union’s Horizon 2022 research and innovation programme
(Grant agreement No. 101076233, "PREXTREME"). 
Views and opinions expressed are however those of the authors only and do not necessarily reflect those of the European Union or the European Research Council Executive Agency. Neither the European Union nor the granting authority can be held responsible for them.

The TDDFT calculations were partly carried out at the Norddeutscher Verbund f\"ur Hoch- und H\"ochstleistungsrechnen (HLRN) under grant mvp00024, and on a Bull Cluster at the Center for Information Services and High Performance Computing (ZIH) at Technische Universit\"at Dresden.

\end{acknowledgements}

%

\end{document}


\title{\underline{Supplemental Material:} Ultrahigh Resolution X-ray Thomson Scattering Measurements at the European XFEL
}

\author{Thomas Gawne}
\email{t.gawne@hzdr.de}

\affiliation{Center for Advanced Systems Understanding (CASUS), D-02826 G\"orlitz, Germany}
\affiliation{Helmholtz-Zentrum Dresden-Rossendorf (HZDR), D-01328 Dresden, Germany}

\author{Zhandos A.~Moldabekov}

\affiliation{Center for Advanced Systems Understanding (CASUS), D-02826 G\"orlitz, Germany}
\affiliation{Helmholtz-Zentrum Dresden-Rossendorf (HZDR), D-01328 Dresden, Germany}

\author{Oliver~S.~Humphries}
\affiliation{European XFEL, D-22869 Schenefeld, Germany}



\author{Karen Appel}
\affiliation{European XFEL, D-22869 Schenefeld, Germany}

\author{Carsten Baehtz}
\affiliation{Helmholtz-Zentrum Dresden-Rossendorf (HZDR), D-01328 Dresden, Germany}

\author{Victorien Bouffetier}
\affiliation{European XFEL, D-22869 Schenefeld, Germany}

\author{Erik Brambrink}
\affiliation{European XFEL, D-22869 Schenefeld, Germany}

\author{Attila Cangi}
\affiliation{Center for Advanced Systems Understanding (CASUS), D-02826 G\"orlitz, Germany}
\affiliation{Helmholtz-Zentrum Dresden-Rossendorf (HZDR), D-01328 Dresden, Germany}

\author{Sebastian G\"ode}
\affiliation{European XFEL, D-22869 Schenefeld, Germany}

\author{Zuzana Kon\^opkov\'a}
\affiliation{European XFEL, D-22869 Schenefeld, Germany}

\author{Mikako Makita}
\affiliation{European XFEL, D-22869 Schenefeld, Germany}

\author{Mikhail Mishchenko}
\affiliation{European XFEL, D-22869 Schenefeld, Germany}

\author{Motoaki Nakatsutsumi}
\affiliation{European XFEL, D-22869 Schenefeld, Germany}

\author{Kushal Ramakrishna}
\affiliation{Center for Advanced Systems Understanding (CASUS), D-02826 G\"orlitz, Germany}
\affiliation{Helmholtz-Zentrum Dresden-Rossendorf (HZDR), D-01328 Dresden, Germany}

\author{Lisa Randolph}
\affiliation{European XFEL, D-22869 Schenefeld, Germany}

\author{Sebastian Schwalbe}

\affiliation{Center for Advanced Systems Understanding (CASUS), D-02826 G\"orlitz, Germany}
\affiliation{Helmholtz-Zentrum Dresden-Rossendorf (HZDR), D-01328 Dresden, Germany}

\author{Jan Vorberger}

\affiliation{Helmholtz-Zentrum Dresden-Rossendorf (HZDR), D-01328 Dresden, Germany}

\author{Lennart Wollenweber}
\affiliation{European XFEL, D-22869 Schenefeld, Germany}

\author{Ulf Zastrau}
\affiliation{European XFEL, D-22869 Schenefeld, Germany}

\author{Tobias Dornheim}
\email{t.dornheim@hzdr.de}

\affiliation{Center for Advanced Systems Understanding (CASUS), D-02826 G\"orlitz, Germany}
\affiliation{Helmholtz-Zentrum Dresden-Rossendorf (HZDR), D-01328 Dresden, Germany}

\author{Thomas~R.~Preston}
\email{thomas.preston@xfel.eu}
\affiliation{European XFEL, D-22869 Schenefeld, Germany}

\maketitle

\subsection*{Experimental details}

The XRTS spectra of Al were measured in the following experimental conditions:

\begin{center}
\begin{tabular}{ c | c | c | c | c}
 $\theta$ (deg.)  & q (${\rm \AA^{-1}}$) & $E_{\rm beam}$ ($\mu$J) & Shots & Frames \\
 \hline
 $3.6  \pm 1.4$ & $0.25 \pm 0.098$  & $18.5 \pm 14.8$ & 85,887   &  300  \\ 
 $8.0  \pm 1.4$ & $0.55 \pm 0.097$  & $19.6 \pm 15.6$ & 92,626   &  450  \\  
 $13.6 \pm 1.4$ & $0.92 \pm 0.097$  & $15.5 \pm 13.3$ & 170,166  & 300   \\ 
 $18.6 \pm 1.4$ & $1.26 \pm 0.096$  & $21.8 \pm 17.0$ & 88,626   &  200  \\
 $25.6 \pm 1.4$ & $1.73 \pm 0.095$  & $21.5 \pm 17.3$ & 137,091  &  300  \\
\end{tabular}
\end{center}
Detector frames were collected at a repetition rate of 10~Hz. Each frame contained 20 x-ray pulses at a repetition rate of 2.2~MHz, so that each frame is the total signal from these 20 pulses. The DCA was scanned in Bragg angle in small $\sim100$\,meV steps and data was continuously acquired for a number of frames shown above.

For all measurements, a linear background from moving off the Rowland circle of the DCA crystal when scanning the energy windows was observed and subtracted.  For the two highest scattering angles, there was an additional background due to the camera sitting in a diffraction peak of Al ($105^\circ$). As the DCA has only a small acceptance window, the diffraction peak was removed by measuring its signal on the camera when the DCA window was far from the peak, then subtracting this mean signal from the full spectrum.

To better resolve the spectral features, the spectra were smoothed with a Savitzky-Golay filter~\cite{savitzky1964smoothing} with a window of 10 pixels ($\sim200$\,meV) and a polynomial order of unity. It was found that the plasmon and elastic peaks were largely unaffected by the filtering as they were already well-resolved by the large number of shots taken at each window position. Therefore, the main effect of the filtering is to smooth the surrounding background noise. The positions of the peaks were then determined by fitting a Voigt profile (\textsc{scipy.special.voigt\_profile} from the SciPy package for Python~\cite{2020SciPy-NMeth}) to the peaks and taking the maximum.

The DCA spectrometer was calibrated using the Co K$_{\beta}$ emission and the position of the quasi-elastic scattering. This was established accurately by tuning the seeded beam energy to cross the Co K-edge and observing when the intensity of the elastically scattered photons plateaued relative to the Co K-shell emission recorded by a spectrometer~\cite{Preston2020} viewing the upstream side of a Co foil. Due to the finite step of moving the seed, there is a small uncertainty in the exact point the K-edge is crossed, leading to an uncertainty in the energy dispersion of the pixels. The energy dispersion was determined to be $22.54 \pm 0.15$~meV/pixel. This leads to a systematic uncertainty in the peak positions of $\pm 0.10-0.16$~eV
(represented as the uncertainty bars along the energy axis in Fig.~3 in the main text),
with peaks at larger energy loss having a larger uncertainty due to the accumulation of the dispersion uncertainty over more pixels.

The uncertainty in the fit parameters for the Bohm-Gross relationship, shown in Fig.~2 of the main paper, was determined by finding the extrema of fits that would still pass through all the $q$ uncertainty bars for the first four wavenumbers.

\subsection*{Time-dependent density functional theory: computational details}

The linear-response TDDFT calculations were performed using the GPAW code~\cite{GPAW1, GPAW2, LRT_GPAW1, LRT_GPAW2, ase-paper, ase-paper2} with a primitive fcc cell of experimental lattice parameter $a=4.05 ~{\rm \AA}$~\cite{wyckoff1948crystal}.
In the simulations, we set the energy cutoff to $1000~{\rm eV}$ (with the PAW dataset provided by GPAW) and the exchange-correlation functional was set to the ground-state local density approximation (LDA) by Perdew and Wang \cite{Perdew_Wang}. For linear-response TDDFT calculations, the momentum transfer must be the difference between two $k$-points. The calculations of the dynamic structure factor (DSF) were performed along the [001] crystallographic direction. We note that the DSF of electrons in fcc Al is isotropic at the considered small wavenumbers $q\lesssim 1~{\rm \AA^{-1}}$~\cite{sprosser1989aluminum}.  We used as a $k$-point grid $N_x\times N_y \times N_z$ with $N_x=40$, $N_y=40$, and with $N_z$ being varied between 40 and 50 to get the DSF at the wavenumbers measured in the experiment. The Lorentzian smearing parameter was set to $\eta=0.05~{\rm eV}$.

\subsection*{Wavenumber averaging}


To account for the effect of $q$-broadening in the TDDFT results, five different TDDFT calculations were performed for each central $q$ value: one at the central $q$, and two either side that are equally spaced. For the slit mask used on the DCA, the number of die collecting varies little with $q$ so that each $q$ point in the range is evenly sampled. Therefore, the TDDFT calculations are averaged with equal weighting to produce the $q$-averaged curve.

\begin{figure*}
    \centering
    \includegraphics[width=0.95\textwidth,keepaspectratio]{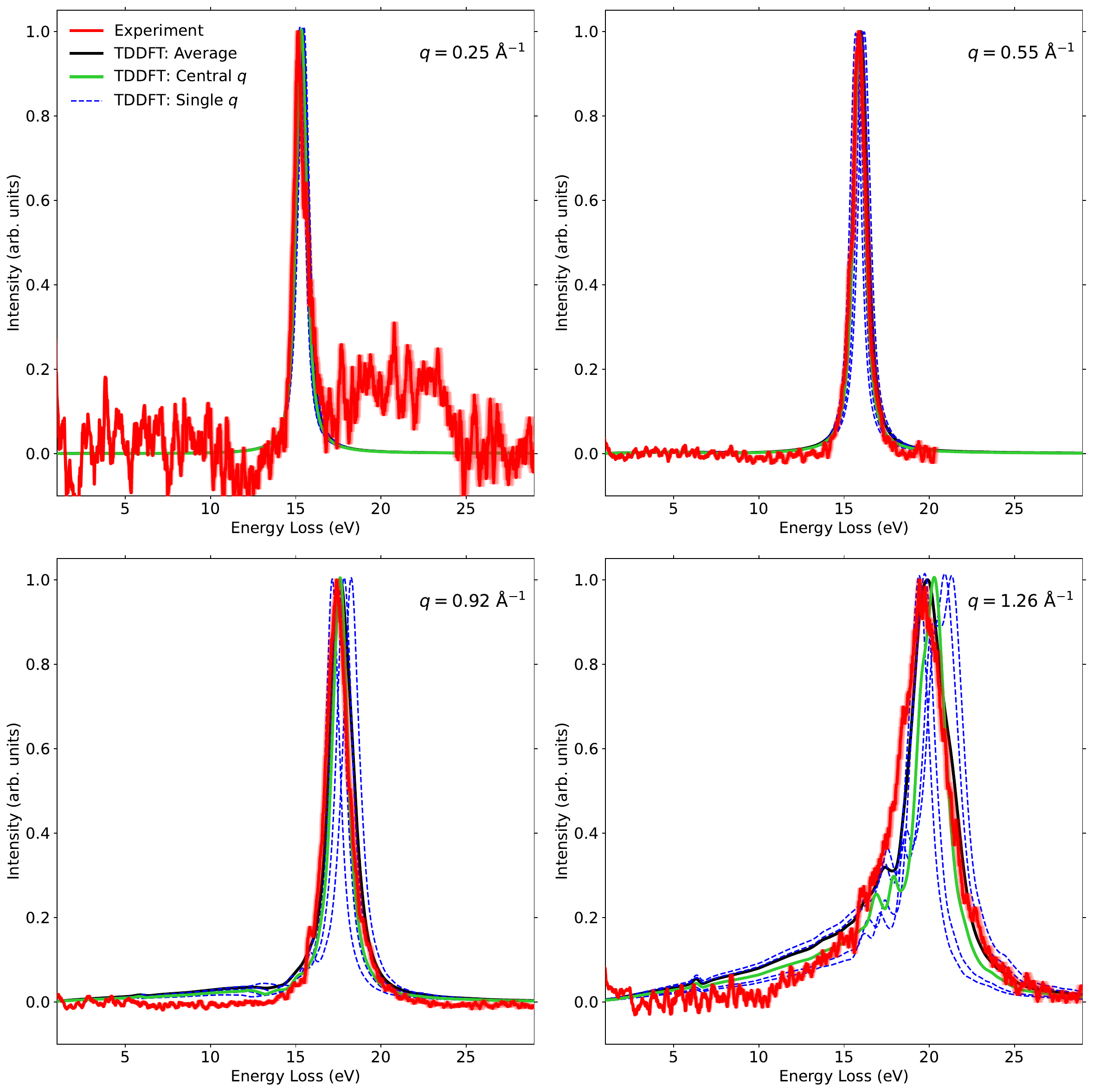}
    \caption{A comparison of TDDFT simulations in the collective regime with XRTS measurements of the plasmon in aluminium at ambient conditions (solid red). The plots include single TDDFT simulations at different $q$-values from within the experimental range (blue dashed), and the average of these 5 individual simulations that are uniformly distributed (green). The individual $q$-values for the different plots are: Top Left: 0.149, 0.199, 0.249, 0.299, 0.348~\,\AA$^{-1}$. Top Right: 0.448, 0.498, 0.537, 0.597, 0.635~\,\AA$^{-1}$. Bottom Left: 0.857, 0.915, 0.932, 0.972, 1.029~\,\AA$^{-1}$. Bottom Right: 1.143, 1.201, 1.258, 1.315, 1.372~\,\AA$^{-1}$.
    }
   \label{fig:Supp1}
\end{figure*}

In Fig.~\ref{fig:Supp1}, we compare our linear response TDDFT simulation results with the XRTS data for the plasmon in aluminium at all four wavenumbers in the collective regime. For the lowest wavenumber $q=0.25\,$\AA$^{-1}$, the spectral weight of the plasmon is comparably small and, as a consequence, the noise level (e.g.~around $E=20\,$eV) significant. The effect of the finite $q$-window in the experimental measurement is small, and the effect of $q$-averaging on the TDDFT results is negligible.
This somewhat changes for $q=0.55\,$\AA$^{-1}$, where some spread of the TDDFT results for different $q$ can be observed; the same holds for $q=0.92\,$\AA$^{-1}$ as discussed in the main text.
Arguably, the most interesting case is given by $q=1.26\,$\AA$^{-1}$, which is located in close proximity to the electron--hole pair continuum, see Fig.~3 of the main text. Here both the experimental measurement and the TDDFT results exhibit a nontrivial structure with a broadened plasmon peak around $E=20\,$eV, and an additional shoulder for smaller energy loss. Moreover, the effect of the $q$-broadening is substantial and affects both the plasmon position and the shape of the spectrum.
Again, we find a noticeable improvement of the $q$-vector averaging over any individual spectrum, and the resulting solid green curve is in good agreement with the experimental data.

\section*{References}
\bibliography{bibliography.bib}